# Improving Axial Resolution of Optical Resolution Photoacoustic Microscopy with Advanced Frequency Domain Eigenspace Based Minimum Variance Beamforming Method

Yu-Hsiang Yu
*Dept. of Electrical Engineering and Brain Research Center*
National Tsing Hua University
Hsinchu, Taiwan
s111061521@m111.nthu.edu.tw

Meng-Lin Li
*Dept. of Electrical Engineering*
Inst. Of Photonic Technologies
Brain Research Center
National Tsing Hua University
Hsinchu, Taiwan
mlli@ee.nthu.edu.tw

*Abstract*—Optical resolution photoacoustic microscopy (OR-PAM) leverages optical focusing and acoustic detection for microscopic optical absorption imaging. Intrinsically it owns high optical lateral resolution and poor acoustic axial resolution. Such anisometric resolution hinders good 3-D visualization; thus 2-D maximum amplitude projection images are commonly presented in the literature. Since its axial resolution is limited by the bandwidth of acoustic detectors, ultrahigh frequency, and wideband detectors with Wiener deconvolution have been proposed to address this issue. Nonetheless, they also introduce other issues such as severe high-frequency attenuation and limited imaging depth. In this work, we view axial resolution improvement as an axial signal reconstruction problem, and the axial resolution degradation is caused by axial sidelobe interference. We propose an advanced frequency-domain eigenspace-based minimum variance (F-EIBMV) beamforming technique to suppress axial sidelobe interference and noises. This method can simultaneously enhance the axial resolution and contrast of OR-PAM. For a 25-MHz OR-PAM system, the full-width at half-maximum of an axial point spread function decreased significantly from 69.3 μm to 16.89 μm, indicating a significant improvement in axial resolution.

## I. Introduction

Optical resolution photoacoustic microscopy (OR-PAM) utilizes optical focusing to obtain a nice lateral resolution of around 1 to 4 μm. Therefore, OR-PAM is capable of producing detailed images, such as the imaging of normalized total hemoglobin concentration in a mouse ear [1] and blood oxygenation in mouse brain with an intact skull [2]. However, in most literature, the images produced by OR-PAM are usually projection views. It is due to the poor axial acoustic resolution which is decided by the acoustic transducer bandwidth. The axial resolution of OR-PAM is approximately 10 to 100 μm making the structures along the depth direction rough to identify in three-dimensional images. Therefore, we aim to improve axial resolution to yield usable three-dimensional images.

In order to enhance axial resolution, C. Zhang *et al.* [3] use ultrahigh frequency detectors larger than 100 MHz along with the Wiener deconvolution method. D. Cai *et al.* [4] employ dual-view OR-PAM and Richardson-Lucy deconvolution to improve axial and lateral resolution. However, these methods require extra equipment to build the system.

We view the axial resolution improvement as an axial signal reconstruction problem, and the axial resolution degradation is caused by axial sidelobe interference. In order to reconstruct the axial signal with higher axial resolution, we need to suppress axial sidelobe interference. In array beamforming, there have been numerous studies dedicated to resolving the sidelobe suppression problem. Our objective is to leverage advanced beamforming techniques to effectively suppress axial sidelobe interference, thereby enabling us to reconstruct axial signals with significantly improved axial resolution.

In addition to sidelobe interference, electronic noise is also a contributing factor. To address these problems, we draw inspiration from eigenspace-based minimum variance beamformer (EIBMV) [5]. The idea is to utilize the eigenstructure of the covariance matrix to separate axial mainlobe contribution from sidelobe interference and noises.

This adaptation of EIBMV to the frequency domain enables us to reconstruct the time domain signal of OR-PAM. Our approach is called frequency domain eigenspace-based minimum variance reconstruction (F-EIBMV).

## II. Materials and Methods

### A. Inverse Discrete Fourier Transform (IDFT)

The OR-PAM signal that we want to reconstruct is a time-domain signal, and the most intuitive way to reconstruct the signal is through inverse discrete Fourier transform as follows.

$$x[n] = \frac{1}{N}\sum_{k=0}^{N-1} X[k]e^{\frac{j2\pi kn}{N}} \quad (1)$$

We can use matrix form to represent IDFT equation as in Eq. (2).

$$x[n] = \vec{W}^H \vec{X'} \quad (2)$$

$$\vec{W} = \frac{1}{N}[1\ 1\cdots 1]^T \quad (3)$$

$$\vec{X'} = [X[0]e^{\frac{j2\pi 0n}{N}}\ X[1]e^{\frac{j2\pi 1n}{N}}\cdots X[N-1]e^{\frac{j2\pi(N-1)n}{N}}]^T \quad (4)$$

$\vec{W}$ is the weight vector which can be viewed as apodization in array beamformation. For standard IDFT the apodization is uniform. $\vec{X'}$ is a vector representing frequency domain signals after phase compensation. The main idea is to manipulate the apodization of IDFT to suppress axial sidelobe interference and noises.

### B. Inverse Discrete Fourier Transform Matrix

Fig. 1(a) is a time domain axial signal without axial sidelobe interference. Fig. 1(b) is the corresponding IDFT matrix, and there is a white horizontal line located at n = 50 which is also the position of time domain axial signal. Fig. 1(c) is the signal value of each frequency at n=50 which shows high spectral coherence.

On the other hand, Fig. 2(a) is a time domain axial signal with axial sidelobe interference. There are 3 gray lines appear on Fig. 2(b). If we again plot the signal value at n = 50, from Fig. 2(c), we can observe that it fluctuates. It is because the original axial signal is affected by the interference. Fig. 2(c) can be viewed as a DC term along with interference, and we want to add apodization to suppress such interference.

### C. Frequency Domain Minimum Variance Method

The way to suppress sidelobe has been exhibited in array beamformation method. One of the most commonly used is the minimum variance beamformer. This is why we adapt minimum variance beamforming technique into the frequency domain.

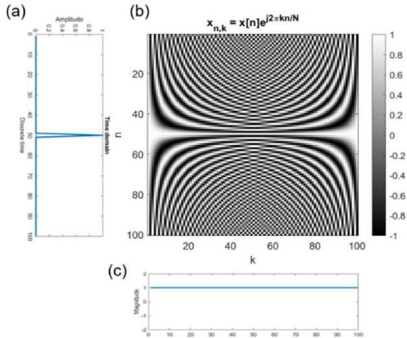

Fig. 1. Axail signal without axial sidelobe interference.

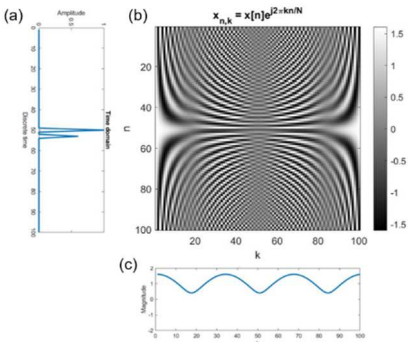

Fig. 2. Axail signal with axial sidelobe interference.

The weight of the frequency domain minimum variance method is calculated by minimizing the power, as follows.

$$\begin{aligned}\vec{W_{MV}} &= \arg\min_{\omega} E\{|x[n]|^2\} \\ &= \arg\min_{\omega} E\left\{|\vec{W}^H \vec{X'}|^2\right\} \\ &= \arg\min_{\omega} \vec{W}^H E\left\{\vec{X'}\vec{X'}^H\right\}\vec{W} \\ &= \arg\min_{\omega} \vec{W}^H R \vec{W}, \text{ subject to } \vec{W}^H \vec{d} = 1\end{aligned} \quad (5)$$

The solution to Eq. (5) is

$$\vec{W_{MV}} = \frac{R^{-1}\vec{d}}{\vec{d}^H R^{-1}\vec{d}} \quad (6)$$

The calculation is basically identical to the original minimum variance array beamformer. However, the difference in the frequency domain minimum variance is that the covariance matrix R is calculated using the spectrum after phase compensation.

### D. Frequency Domain Eigenspace Based Minimum Variance Reconstruction (F-EIBMV)

In addition to axial sidelobe interference, there are electronic noises in the OR-PAM system. We hope to use eigen decomposition to separate axial mainlobe signal and noises.

Learning from eigenspace-based minimum variance beamformer [5]. We also perform eigen decomposition on the covariance matrix R.

$$R = V\Lambda^{-1}V^H \quad (7)$$

, where, the eigenvalues are $\Lambda = \text{diag}[\lambda_1, \lambda_2, \cdots, \lambda_{N-1}]$, $\lambda_1 \geq \lambda_2 \geq \cdots \geq \lambda_{N-1}$, and $V = [\mathbf{v}_1, \mathbf{v}_2, \cdots, \mathbf{v}_{N-1}]$ are the eigenvectors. Then, using eigenvectors with larger eigenvalues to construct signal subspace $E_s$.

$$E_s = [\mathbf{v}_1, \cdots, \mathbf{v}_{\text{Num}}] \quad (8)$$

The Num is the number of eigenvectors that contain axial mainlobe. Therefore, by choosing sufficient number of eigenvectors, we are able to reconstruct axial signal without axial sidelobe interference and noises.

Next, projecting the original frequency domain minimum variance weight $\vec{W_{MV}}$ onto signal subspace yielding the desired frequency domain eigenspace based minimum variance weight $\vec{W_{EIBMV}}$.

$$\vec{W_{EIBMV}} = E_s E_s^H \vec{W_{MV}} \quad (9)$$

Finally, replace the original uniform weight in (2) with frequency domain eigenspace based minimum variance weight.

$$x[n] = \vec{W_{EIBMV}}^H \vec{X'} \quad (10)$$

By doing so, we can reconstruct OR-PAM's time domain signal with suppressed axial sidelobe interference and noises which can enhance axial resolution and imaging contrast.

### E. Optical resolution photoacoustic microscopy (OR-PAM)

Fig. 3. is the setup of our 25 MHz OR-PAM system. The 532 nm pulsed laser is guided through optical fibers to a Galvo for two-dimensional laser scanning and subsequently focused using an objective lens. The focused beam is directed through a 3D-printed photoacoustic beam combiner and excites the sample. When the laser's energy is absorbed by the sample, the local spot experiences thermal elastic expansion and

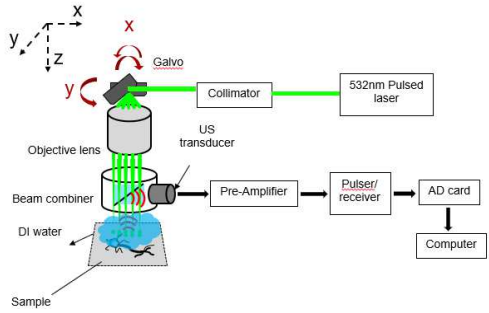

Fig. 3. The setup of the used 25 MHz OR-PAM

generates a pressure wave (or acoustic wave) that propagates outward. A glass slide set at a 45-degree angle positioned at the center of beam combiner, enabling laser transmission while reflecting the photoacoustic signal onto a 25 MHz ultrasound transducer.

The photoacoustic signal is converted into electrical signals stored in a computer for imaging. The photoacoustic signal received by the transducer underwent initial pre-amplification through a low-noise amplifier. Subsequently, it was filtered and amplified once more using a pulser/receiver. Finally, the signal was digitized by a data acquisition card operating at a sampling rate of 200 MS/s and stored in a computer.

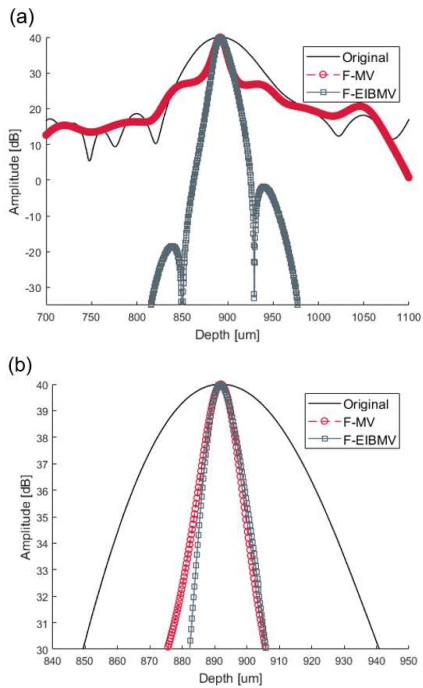

Fig. 4. Axail point spread function. (b) is the zoom in of (a)

## III. RESULTS

In order to see the axial resolution improvement, we measure the axial point spread function of a 25 MHz OR-PAM system by directing the laser onto an airforce target. Because airforce target is a thin film, so we can extract an A-line to evaluate the full width at half-maximum and estimate axial resolution.

In Fig. 4, with the frequency domain minimum variance (F-MV) method, the axial sidelobe interference is suppressed, resulting in an improvement in axial resolution. Moreover, by applying the proposed method, the electronic noises can be further reduced by at least 20 dB. The overall image contrast will be enhanced because of the reduced noise levels.

The full-width at half-maximum (FWHM) of the original axial point spread function (PSF) is 69.3 μm, and that of the axial PSF after applying F-MV is 18.38 μm. For the proposed method, the FWHM is 16.89 μm which is approximately 4 times thinner than the original axial PSF.

### ACKNOWLEDGMENTS

The authors appreciate the support of National Science and Technology Council, Taiwan (MOST 110-2221-E-007-011 -MY3, NSTC 111-2321-B-002-016- and NSTC 112-2321-B-002-025-) and Brain Research Center, NTHU, under the Higher Education Sprout Project, funded by the Ministry of Education, Taiwan.